\documentclass[preprint,12pt]{emulateapj}

\usepackage{natbib}

\usepackage{graphicx}
\usepackage[space]{grffile}
\usepackage{latexsym}
\usepackage{amsfonts,amsmath,amssymb}
\usepackage{url}
\usepackage[utf8]{inputenc}
\usepackage{fancyref}
\usepackage{hyperref}
\hypersetup{colorlinks=false,pdfborder={0 0 0},}
\usepackage{textcomp}
\usepackage{longtable}
\usepackage{multirow,booktabs}
\usepackage{subfigure}
\usepackage{tablefootnote}

\newcommand{\hd}{{HIP~41378}\xspace}

\usepackage{xspace}
\usepackage{amsmath}
\usepackage{xcolor}

\shorttitle{Revisiting HIP41378}
\shortauthors{Berardo et al.}

\begin{document}


\title{Revisiting the HIP41378 system with \textit{K2} and \textit{Spitzer}}


\author{
David Berardo\altaffilmark{1,2,,$\dagger$},
Ian J.\ M.\ Crossfield\altaffilmark{1},
Michael Werner\altaffilmark{3},
Erik Petigura\altaffilmark{4},
Jessie Christiansen\altaffilmark{5},
David R. Ciardi\altaffilmark{5},
Courtney Dressing\altaffilmark{6},
Benjamin J. Fulton\altaffilmark{4},
Varoujan Gorjian\altaffilmark{3},
Thomas P. Greene\altaffilmark{7},
Kevin Hardegree-Ullman\altaffilmark{5},
Stephen R. Kane\altaffilmark{9},
John Livingston\altaffilmark{10},
Farisa Morales\altaffilmark{3},
Joshua E. Schlieder\altaffilmark{11}
}

\altaffiltext{1}{Department of Physics, and Kavli Institute for Astrophysics and Space Research, Massachusetts Institute of Technology, Cambridge, MA 02139, USA}
\altaffiltext{2}{berardo@mit.edu}

\altaffiltext{3}{Jet Propulsion Laboratory, California Institute of Technology, Pasadena, CA 91109, USA}

\altaffiltext{4}{1Division of Geological and Planetary Sciences, California Institute of Technology, Pasadena, CA 91125, USA}

\altaffiltext{5}{NASA Exoplanet Science Institute, California Institute of
Technology, M/S 100-22, 770 S. Wilson Ave, Pasadena, CA, USA}

\altaffiltext{6}{Astronomy Department, University of California, Berkeley, CA 94720, USA}
\altaffiltext{7}{NASA Ames Research Center, Space Science and Astrobiology Division, MS 245-6, Moffett Field, CA 94035}

\altaffiltext{8}{Department of Physics and Astronomy, University of Toledo, 2801 W. Bancroft Street, Toledo, OH 43606, USA}

\altaffiltext{9}{University of California Riverside, Department of Earth Sciences, Riverside, CA 92521, USA}

\altaffiltext{10}{Department of Astronomy, University of Tokyo, 7-3-1 Hongo, Bunkyo-ky, Tokyo 113-0033, Japan}
\altaffiltext{11}{Exoplanets and Stellar Astrophysics Laboratory, NASA Goddard Space Flight Center, Mail Code 667, 8800 Greenbelt Rd, Greenbelt, MD 20771, USA}

\altaffiltext{$\dagger$}{NSERC Postgraduate Scholarship - Doctoral}

\begin{abstract}

We present new observations of the multi-planet system HIP 41378, a bright star ($V$ = 8.9, $K_s$ = 7.7) with five known transiting planets.  Previous \textit{K2} observations showed multiple transits of two Neptune-sized bodies and single transits of three larger planets ($R_P= 0.33R_J, 0.47R_J, 0.88 R_J$). \textit{K2} recently observed the system again in Campaign 18 (C18). We observe one new transit each of two of the larger planets d/f, giving maximal orbital periods of 1114/1084 days, as well as integer divisions of these values down to a lower limit of about 50 days. We use all available photometry to determine the eccentricity distributions of HIP41378 d \& f, finding that periods $\lesssim$300 days require non-zero eccentricity. We check for overlapping orbits of planets d and f to constrain their mutual periods, finding that short periods (P $<$ 300 days) for planet f are disfavoured. We also observe transits of planets b and c with Spitzer/IRAC, which we combine with the \textit{K2} observations to search for transit timing variations (TTVs). We find a linear ephemeris for planet b, but see a significant TTV signal for planet c. The ability to recover the two smaller planets with \textit{Spitzer} shows that this fascinating system will continue to be detectable with \textit{Spitzer}, \textit{CHEOPS}, \textit{TESS}, and other observatories, allowing us to precisely determine the periods of d and f, characterize the TTVs of planet c, recover the transits of planet e, and further enhance our view of this remarkable dynamical laboratory.

\end{abstract}


\bibliographystyle{apj}
\section{Introduction}
\setcounter{footnote}{0}
Multi-planetary systems are just one of the many exciting discoveries that NASA's \textit{Kepler} and \textit{K2} missions have produced since the spacecraft's launch in 2009. These systems allow us to probe details regarding the formation, stability, and general structure of exoplanets, providing crucial data to motivate theories of exoplanet dynamics \citep[e.g.,][]{becker:2015, weiss:2018a}. Although the \textit{K2} mission is winding down, as we enter the next generation of exoplanet missions (\textit{TESS}, \textit{CHEOPS}, and eventually \textit{JWST} and \textit{ARIEL}), \textit{K2} has proven its usefulness yet again with new observations of the multiplanet HIP41378 system\footnote{RA: 08h26m27.85s, DEC: +10d04m49.4s}, which it previously observed during Campaign 5 (C5) \citep{vanderburg:2016c}. 

The initial observation revealed a rich system of two shorter-period planets and three single transit events, which were statistically significant as planets. As is often the case with such systems, additional data were needed to refine the orbital and physical properties of these outer planets and this was recently provided by \textit{K2} during Campaign 18 (C18), which took both long and short cadence observations of HIP~41378. This system is not only one of a handful of known stars hosting five planets, but is also the second brightest such system, with the host star having a V band magnitude of 8.9 and K magnitude of 7.7  - beaten only by the 55 Cancri system \citep{fischer:2008} - making it a compelling target for future characterization if the periods of the larger planets can be precisely determined.

In Sec.~\ref{sec:obs} we discuss the various observations and analysis methods we use to further characterize the system. In Sec.~\ref{sec:stellar} we provide updated stellar parameters for the host star based on \textit{Gaia} data. Sec.~\ref{sec:photoeccentric} discusses the techniques and results of our dynamical study of the system, including eccentricity estimates. Finally in Sec.~\ref{sec:conclusions} we summarize our results and discuss the potential for future observations.

\newpage
\section{Photometric Observations and Analysis}
\label{sec:obs}
Below we describe our time-series photometry analysis of \hd, which includes photometry from \textit{K2} (Sec.~\ref{sec:k2obs}), from \textit{Spitzer} (Sec.~\ref{sec:spitzer}), and a joint analysis of data sets from both telescopes (Sec.~\ref{sec:joint}).

\subsection{K2}
\label{sec:k2obs}
\hd was originally observed by the \textit{Kepler} space telescope during Campaign 5 of the \textit{K2} mission for approximately 75 days. The system was then observed again during Campaign 18 for approximately 50 days\footnote{While there was also partial overlap between the fields of Campaigns 5, 16 and 18, \hd\ was not observed in C16.}$^{,}$\footnote{Long cadence observations proposed for in C18 GO programs 3, 6, 27, 36, 47, 49, 65, 67, 901}. Additionally, since the system was known to host planets, short cadence (1 minute) photometry was collected during C18\footnote{Short cadence observations proposed for in C18 GO programs 6,27,36,47}. The C5 data spans from BJD$_{TDB}$ = 2457140.5 to 2457214.4 and is composed of 3378 frames, corresponding to observations every 30 minutes (with frames removed for thruster firing and other data quality flags). The C18 data spans BJD$_{TDB}$ =  2458251.5 to 2458302.4 and consists of 2195 frames for the long cadence data and 60000 frames for the short cadence data, again with frames removed due to quality issues. Thus there is a gap of approximately 1037 days between the end of C5 and the beginning of C18.

\begin{figure*}[bt!]
\begin{center}
\includegraphics[width=6.5in,angle=0]{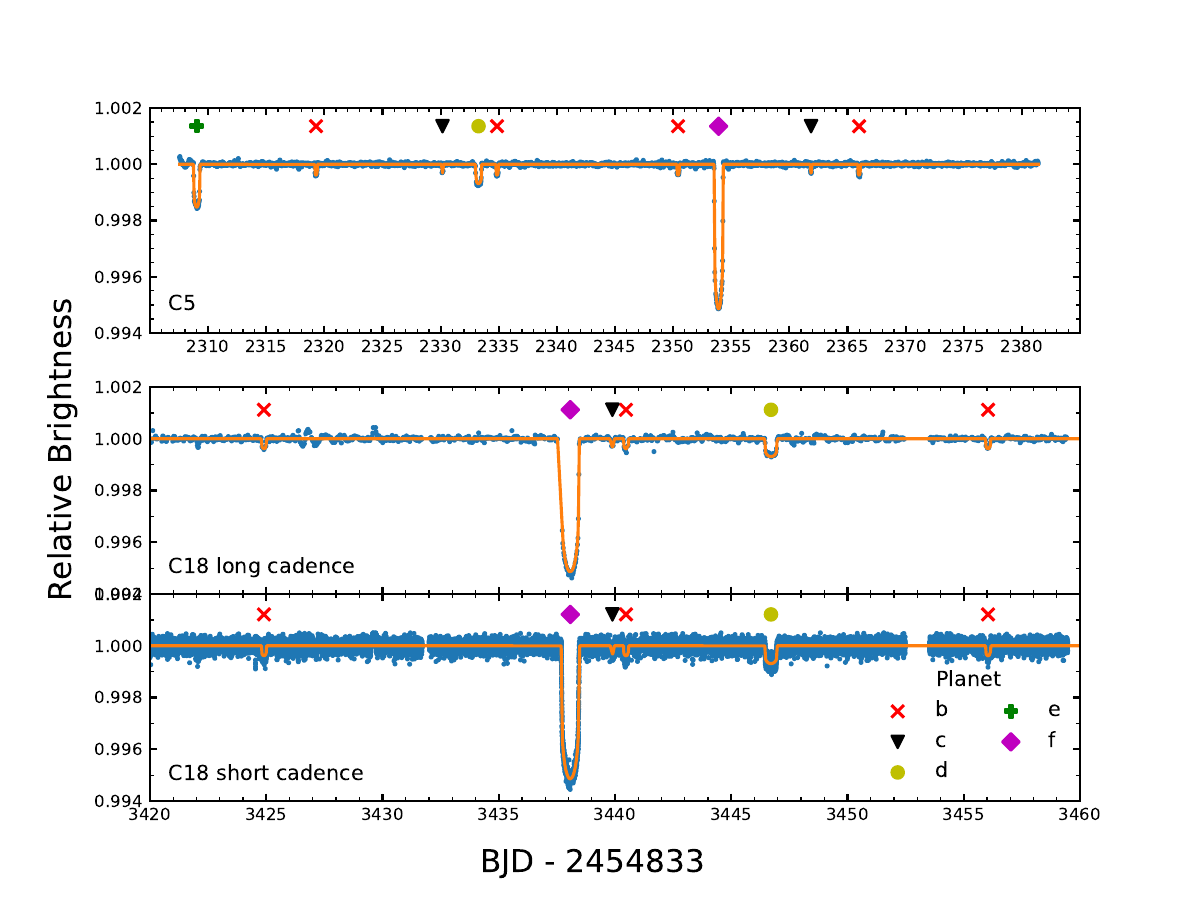}
\caption{\label{fig:k2 lightcurves} Light curves of \hd\ across all three observation (in blue) with transits of each planet highlighted. The orange line represents the best fit model.}
\end{center}
\end{figure*}

\textbf{In the analysis that follows we use the fully detrended C5 lightcurve provided by A.~Vanderburg \citep{vanderburg:2016c}. The calibrated C18 short and long cadence data are downloaded from MAST as target pixel files. These are then converted into lightcurves by performing simple aperture photometry, using a circular aperture centered on the 'center of light' of each image to measure the stellar flux. The center of light is determined by taking the weighted mean of the flux of each pixel in the image. We then detrend both the long and short cadence lightcurves following the methods outlined in \cite{vanderburg:2014}.} Low frequency variations in each lightcurve are removed by first masking out points associated with transits, and then fitting a basis spline and dividing out the best fit to produce flattened light curves. Additionally, we trimmed the data  to include only points within two transit durations from an expected transit center, to reduce analysis run times (Figure \ref{fig:k2 lightcurves} shows the full short cadence lightcurve.). This is done in order to fit for individual planets without interference from the transit signals of the other planets in the system. This process produces 15 lightcurves, corresponding to three observations times five planets. Before trimming, we also check the lightcurves for signs of planet e, and while there additional transit-like signals in the lightcurve, none of them agree with the depth or duration of the known planets in the system and are likely due to detrending issues.



\begin{figure*}[bt!]
\begin{center}
\includegraphics[width=7.5in,height=5in,angle=0]{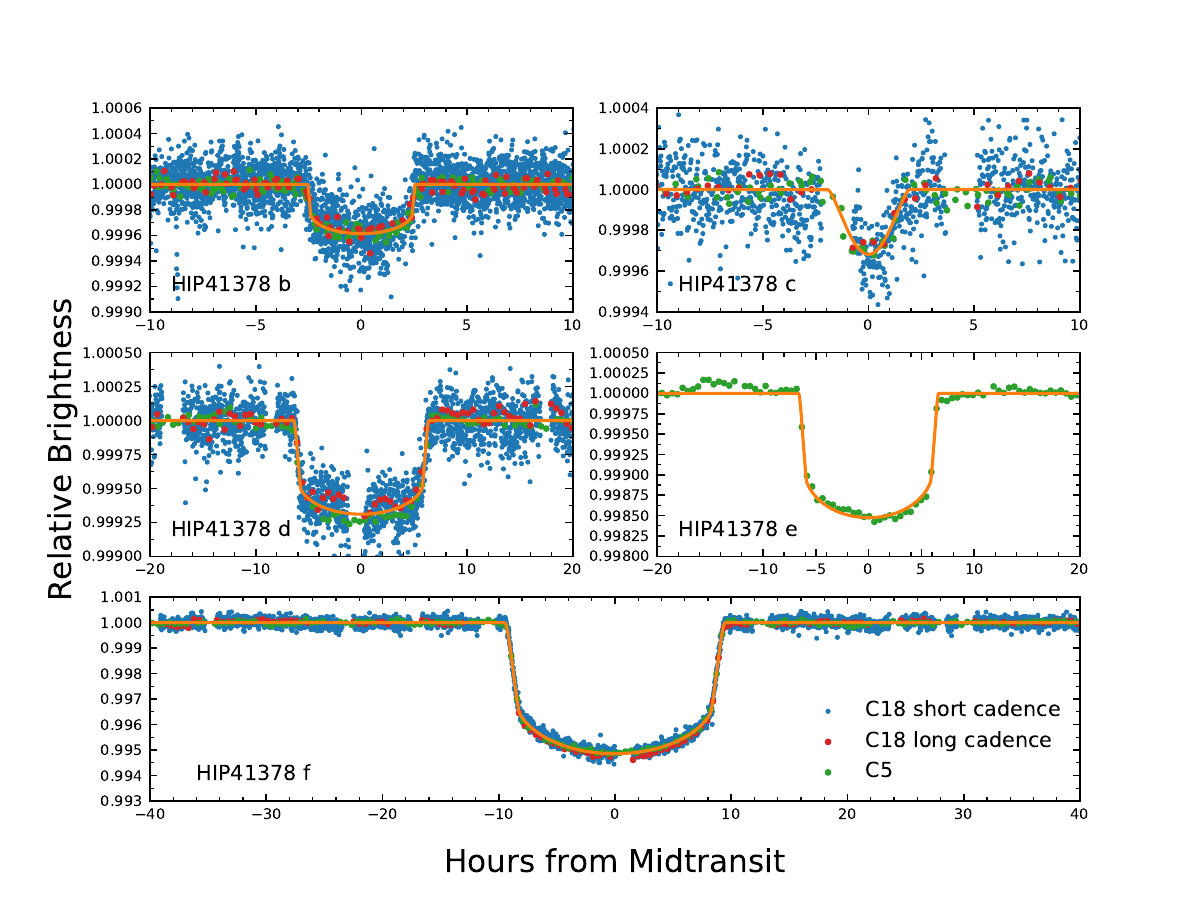}
\caption{\label{fig:individual transits} Transits of all five planets of \hd, showing \textit{K2} data from C5 (green points) and C18 short and long cadence (blue and red points), along with our best-fit transit models (solid orange line). Planet e was not observed to transit during C18, and so we only show C5 data for it.}
\end{center}
\end{figure*}

We derive a best fit lightcurve model by fitting for each planet individually, using the \texttt{batman}\footnote{https://github.com/lkreidberg/batman} \citep{kreidberg:2015b} and \texttt{emcee}\footnote{http://dfm.io/emcee/current/} \citep{foreman-mackey:2012} Python packages to perform an MCMC analysis. \textbf{When calculating lightcurves with \texttt{batman}, we divide the lightcurve into 30 minute intervals and then average over 10 points within each interval. This is done to simulate the 30 minute cadence of the K2 data. For 1 minute cadence, it was found that averaging over 1 minute intervals changed the lightcurve at a level well below the scatter of the data, so averaging was not done.} We evolve 150 walkers  for 20,000 burn-in steps, followed by an additional 20,000 steps which are used to estimate the posterior values of the fitted parameters. These parameters are the center of transit $t_{0}$, orbital period $p$, scaled planet radius $r_p / r_s$ (\textbf{where $r_s$ is the radius of the host star)}, scaled semi-major axis $a / r_s$, orbital inclination $i$, and two limb darkening parameters for a quadratic limb darkening model $q_1, q_2$ \citep{kipping:2013}. Additionally, the scatter $\sigma$ of each lightcurve is left as a free parameter during the fit, producing three additional parameters (one for each observation). Thus the likelihood being maximized has the form:

\begin{equation}
\mathrm{Ln}\mathcal{L} = -\frac{1}{2}\sum_{i} \frac{(\mathrm{flux_i} - \mathrm{model_i})^2}{\sigma_i^2} - 2\mathrm{Ln}(\sigma_i)
\end{equation}

where the index i runs over the three observations, flux is the observed lightcurve, and the model is the calculated lightcurve  given a set of orbital parameters. \textbf{For all parameters except the limb darkening coefficients, we use flat priors. We also use a flat prior for the limb darkening coefficients when fitting for planet f. The high signal to noise of planet f produces the tightest constraints on these parameters. For planets b through e, we then use the values of q1 and q2 found for planet f as gaussian priors when running the MCMC.} The resulting parameters are given in Table~\ref{tab:k2 fit params}, \textbf{where we quote the median value of the MCMC posterior distribution with 68\% confidence intervals}. The best fit models are shown in Figure \ref{fig:k2 lightcurves}\footnote{For planets d and f we show the fit results assuming the maximal period, although other periods are possible as discussed in Sec.~\ref{sec:photoeccentric}}, \textbf{and in Figure \ref{fig:individual transits} we show the individual lightcurves for each planet, phase folded on their respective periods}. In each case, the posterior value for the scatter is consistent with the out of transit standard deviation of the lightcurves.

\subsection{Spitzer}
\label{sec:spitzer}

We also observed \hd\ using the 4.5$\mu m$ channel on Spitzer's IRAC camera as part of observing programs 11026 centered on BJD\_UTC 2457606.932 and 13052 centered on BJD\_UTC 2457790.680  (PI Werner). The first observation coincides with an expected transit of HIP41378 c while the second corresponds to an expected transit of HIP41378 b.  

We downloaded data from the \textit{Spitzer} Heritage 
Archive\footnote{http://sha.ipac.caltech.edu/applications/Spitzer/SHA/} and processed it into lightcurves as follows. First, we used a median filter with a span of 10 frames  to remove anomalous pixels (flux values $> 4\sigma$ from the median) due to cosmic rays and other effects. The centroid of each frame is then calculated in two ways, once by fitting a two dimensional Gaussian brightness profile, and again by calculating the center of light:
\begin{equation}
x_c = \frac{\sum_i f_i x_i}{\sum f_i}, y_c = \frac{\sum_i f_i y_i}{\sum f_i}
\end{equation}
where $f_i$ is the flux of the i$^{th}$ column and $x_i$ is the x-position for of the i$^{th}$ column (similarly for $y$ and the rows). For each frame, we also calculated the background level by taking the flux in a 10$\times$10 square in each corner of the image, fitting a Gaussian to the distribution of flux values, and taking the mean of the Gaussian to be the background level.

Light curves are then computed by summing up the flux in a circular aperture around the centroid and subtracting the appropriate amount of background flux, using the \texttt{photutils} \citep{bradley:2018} python package to account for partial pixels. We do this for apertures whose radii span from 1.8 to 3.4 pixels in 0.2 pixel increments, producing lightcurves for each combination of centroid method and aperture radius. For each of these, we determine a best fit systematics model using the Pixel Level De-correlation (PLD) technique \citep{deming:2015}. This method attempts to correct for the varying response of the pixels as the centroid moves around the CCD. Despite centroid motions of only about a tenth of a pixel, the magnitude of the intrapixel sensitivity, combined with the shallow depths of the transits (100's of ppm) requires detrending of this effect in order to recover the transits. 

We model the total flux as
\begin{equation}
S = \sum_i c_i f_i + D E(t) + ht + gt^2
\end{equation}
where D is the transit depth, E(t) is the transit model, $f_i$ is the flux of the i'th pixel, $c_i$ are coefficients correcting for the varying response of the pixels, and h and g are parameters used to model a quadratic time ramp.
We perform a $\chi^2$ minimization for each lightcurve to determine the best fit parameters, and use the quality of the fits to determine which lightcurve to ultimately use. This is done by binning the residuals of the fit, plotting the standard deviation versus bin factor, and choosing the one which has the closest slope to -0.5 (in log space), indicating the lowest amount of correlated noise. In addition to choosing the best lightcurve, we also bin down the data and see what effect this has on the results as well. This procedure ends up selecting a 2D-Gaussian fit for centroiding, an aperture radius of 2.4 pixels, and a bin size of 200 points per bin.

\begin{figure*}[ht!!]
\begin{center}
\includegraphics[width=7.5in,angle=0]{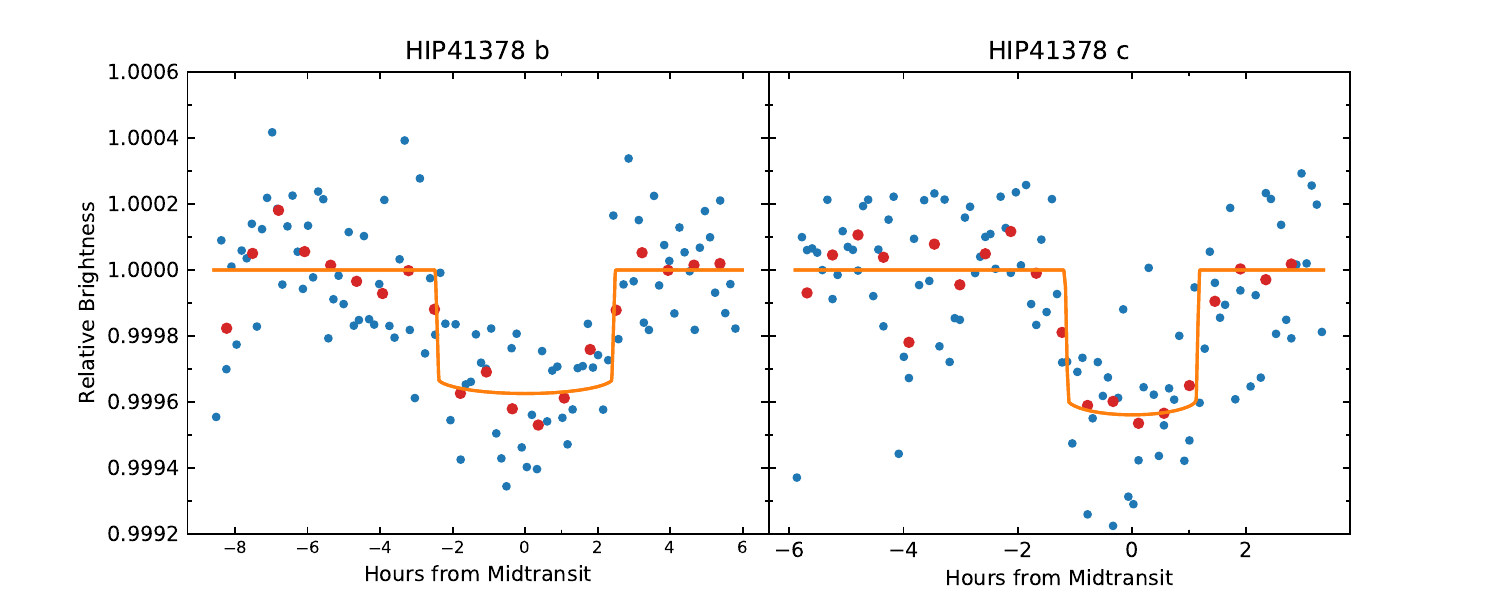}
\caption{\label{fig:spitzer transits} \textit{Spitzer} photometry of HIP~41378b (left) and~c (right). Blue dots show the (binned) photometry after removing systematic effects, red dots show the photometry binned by an additional factor of five, and the solid orange line shows the best fit transit models.}
\end{center}
\end{figure*}

Once we have chosen the best lightcurve for each observation, as for the \textit{K2} data we run an MCMC chain in order to obtain posterior probability distributions and determine the errors for each parameter. The values being fit during the MCMC are the PLD pixel coefficients, the two time ramp parameters, the center of transit, the transit depth, as well as the orbital inclination and semi-major axis. The best fit transit signals are shown in Figure~\ref{fig:spitzer transits} and the values are listed in Table~\ref{tab:spitzerfitparams}.

We also performed analyses with two independent implementations of PLD, fitting the \textit{Spitzer} data by itself (Hardegree-Ullman et al., in preparation) and also simultaneously with the \textit{K2} data (Livingston et al., in review), and the resulting parameter estimates were consistent. \textbf{We find that for both planets, the values of semi-major axis and depth are consistent within the quoted errors between the \textit{Spitzer} and \textit{K2} values.}

\subsection{Joint K2+Spitzer Analysis}
\label{sec:joint}

\begin{figure}[ht!]
\begin{center}
\includegraphics[width=3.5in,angle=0]{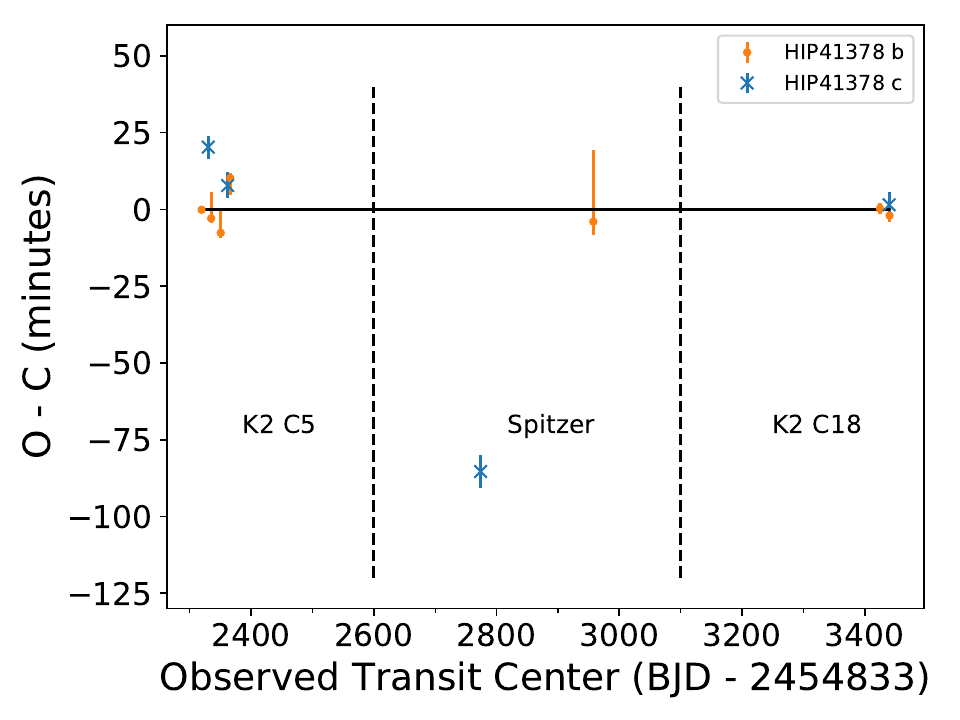}
\caption{\label{fig:ttvs}Transit timing variations plot for HIP41378 b and c. Here we show the deviation from a linear ephemeris as a function of measured transit center for the two inner short period planets of the system. The dashed lines separate the three observations. 
}
\end{center}
\end{figure}

Combining the \textit{K2} and \textit{Spitzer} observations provides a total of 8 transits of HIP41378~b and 4 transit of HIP41378~c, which allows us to check for transit timing variations (TTVs) that could indicate the presence of other non-transiting bodies and/or constrain the planets' masses. For both planets b and c, we keep fixed all of the best-fit parameters described in Sec.~\ref{sec:k2obs} and re-fit each transit individually across the C5, C18 short cadence, and \textit{Spitzer} data, allowing only the transit center to vary. For each planet we then fit a linear ephemeris to their epochs and observed transit times (taking into account their relative uncertainties), and plot the difference between the observed and calculated values in Figure~\ref{fig:ttvs} (these values are also listed in Tables \ref{tab:ttv times b} and \ref{tab:ttv times c}). For planet b we discard the last observation, where we find a large offset in the transit center which we attribute to our detrending of the short cadence C18 data. We feel comfortable discarding this point since we have two other transits of planet b during C18 to establish a long baseline with previous observations. 

For HIP41378~b we find results consistent with a linear ephemeris. For HIP41378 c, we find that the the transit times are inconsistent with a linear ephemeris. While the systematic effects of the \textit{Spitzer} observation make it difficult to obtain precise orbital parameters, as mentioned in Sec.~\ref{sec:spitzer} we have two external independent analyses of the observations which both produce similar TTV signals. We note while the C5 and \textit{Spitzer} observations are broadly consistent with a linear ephemeris, although they predict that the transit of HIP41378 c in C18 should be $\sim 3$ hours from where it is currently measured. Despite larger scatter in the C18 data than the C5 data, we do not believe that the transit center would shift by that amount. Additional transits are required to confirm the TTV signal seen here (see Figure \ref{fig:ttvs} and Sec.~\ref{sec: future obs}).

\section{Stellar Parameters}
\label{sec:stellar}
We derive an updated set of stellar parameters for HIP~41378 for use in our subsequent analysis. \cite{vanderburg:2016c} report $T_\mathrm{eff}=6199 \pm 50$~K using spectroscopic techniques.   We infer $T_\mathrm{eff}=6283 \pm 43$ by comparing the $B-V$, $V-K_s$, and $J-H$ colors to Table 5 of \cite{pecaut:2013} and taking a weighted mean of the individual values from each color.   We use the weighted mean of these two independent temperatures,  along with the Gaia DR2 parallax \citep{gaia:2016a, gaia:2018} and the apparent stellar magnitudes in $Ks$ and $W1$, as input parameters for the \texttt{isochrones} package \citep{morton:2015b} with the MIST tracks \citep{choi:2016}. The computed parameters are $T_\mathrm{eff}=6226 \pm 43$~K, $R_*=1.375 \pm 0.021\ R_\odot$, $M_*= 1.168 \pm 0.072\ M_\odot$, and $d = 106.58 \pm 0.65$~pc.  None of these (except $R_*$) change by more than 1.5$\sigma$ if we instead use the parallax with the magnitudes in $V$, $B$, $J$, $H$, $K_s$, $W1$, and $W2$.  In this second analysis we find $R_*=1.310 \pm 0.016\ R_\odot$, so we take the mean and report an uncertainty that covers both values. Thus our final stellar radius is $1.343 \pm 0.032\ R_\odot$.  Our updated stellar parameters are listed in Table~\ref{tab:stellar}; all are consistent with (but more precise than) those of \cite{vanderburg:2016c}.

We also derive stellar parameters using a high-resolution optical spectrum taken from Keck/HIRES, following the approach of \cite{fulton:2018} . This spectrum implies $T_\mathrm{eff}=6266 \pm 100$, $R_*=1.33 \pm 0.013\ R_\odot$, $M_*= 1.17 \pm 0.030\ M_\odot$, consistent with our analysis above.

Finally, we observe solar-like oscillations in the C18 short cadence data. These could further refine the stellar parameters, but we defer that analysis for a subsequent paper.

\begin{deluxetable}{l l c l}[bt]
\tabletypesize{\scriptsize}
\tablecaption{Updated HIP~41378 Parameters\label{tab:stellar}}
\tablehead{
\colhead{Parameter} & \colhead{Units} & \colhead{Value} & \colhead{Comment}
}
\startdata
$\varpi$ & mas & $ 9.3799 \pm 0.059 $ & \citeauthor{gaia:2018} \\
$R_*$ & $R_\odot$ & $1.343 \pm 0.032$ & This work, Sec.~\ref{sec:stellar}\\
$M_*$ & $M_\odot$ & $1.168 \pm 0.072$ & This work, Sec.~\ref{sec:stellar}\\
$\rho_*$ & g~cm$^{-3}$ & $0.680 \pm 0.064$ & This work, Sec.~\ref{sec:stellar}\\
$T_{eff}$ & K & $6226  \pm 43$ & This work, Sec.~\ref{sec:stellar}
\enddata
\end{deluxetable}


\section{Dynamics}
\label{sec:photoeccentric}
We used the transits of HIP~41378 f and HIP~41378 d to constrain each planet's orbital eccentricity by applying the ``photoeccentric'' formalism of \cite{dawson:2012}, using the same software and approach  as described by \cite{schlieder:2016}. This technique uses knowledge of the true stellar density $\rho_\star$ (calculated using our parameters in Sec~\ref{sec:stellar}), combined with the derived stellar density from a fit assuming zero eccentricity $\rho_{\star,circ}$,

\begin{equation}
\rho_{\star,circ} = \frac{3\pi (a / r_s)^3}{GP^2}
\end{equation}

in order to estimate the eccentricity of the orbit, where $a/r_s$ is the scaled semi-major axis and $P$ is the orbital period.

Since the two transits of HIP41378 d/f have a gap of $\sim$ 1000 days between them, there is a range of allowed periods that would produce the observed signals. The maximal possible period for the two planets, given by the delay between the observed transits, is 1114 days for planet d and 1084 days for planet f. The minimum possible periods are 48 days for planet d and 46 for planet f (shorter periods would have produced additional transits in either C5 or C18). Any fractional value of the longest period is also valid, and so this gives a range of 23 possible periods for both planets, for each of which we perform a photoeccentric analysis\footnote{None of the allowed periods predict transits of planet d or f during our \textit{Spitzer} observations.}. We show the results of the five longest (and most plausible, as described below) periods for each planet in Tables~\ref{tab:photoeccentricf} \& \ref{tab:photoeccentricd}, listing the maximum-likelihood values and 15.8\% and 84.2\% confidence intervals for all parameters. In addition to $e$ and $\omega$, we include the photoeccentric parameter $g$,
\begin{equation}
g(e,\omega) = \frac{1+e \sin \omega}{\sqrt{1-e^2}} = \left( \frac{\rho_{*,circ}}{\rho_{*}} \right) ^{1/3}
\end{equation}
See Fig.~2 of \cite{dawson:2016} for the allowed relationships between $e$ and $\omega$ for various values of $g$. For $\rho_*$, the stellar density, we use the value in Table~\ref{tab:stellar}. For $\rho_{*,circ}$, the density inferred solely from the transit light curve assuming a circular orbit \citep{seager:2003}, we take the posteriors computed directly from our MCMC analyses.

\begin{figure}[bt!]
\begin{center}
\includegraphics[width=3.5in,angle=0]{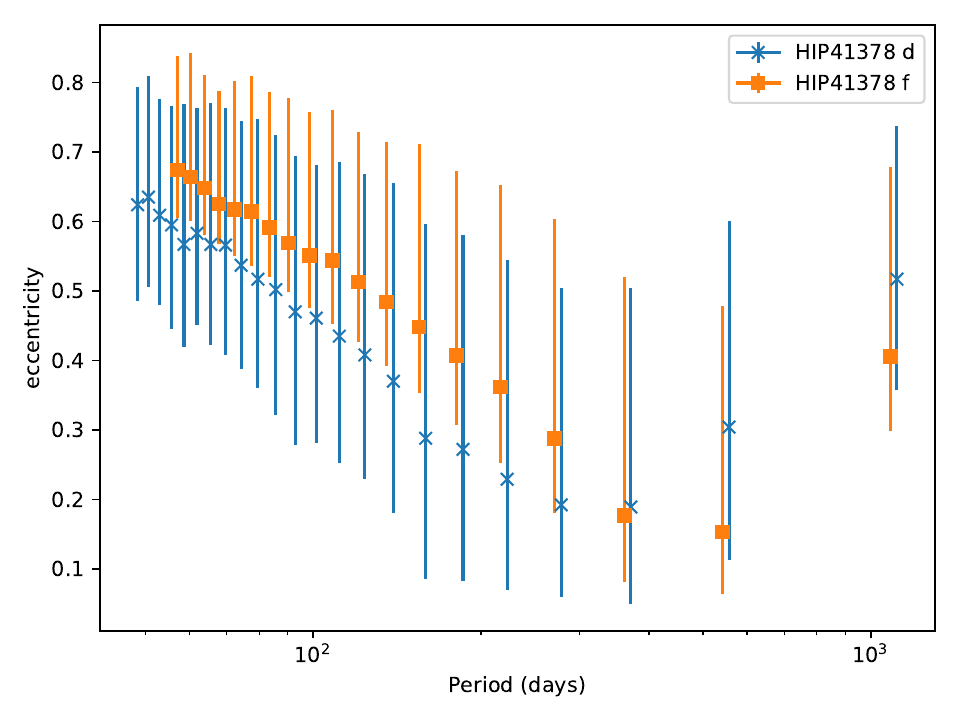}
\caption{\label{fig:eccentricites} Possible eccentricities of HIP~41378d (blue) and~f (orange) from our photoeccentric analyses.  We see a similar decreasing trend in eccentricity for both planets, indicating that lower eccentricities are consistent with longer periods.}
\end{center}
\end{figure}

\begin{figure}[bt!]
\begin{center}
\includegraphics[width=3.5in,angle=0]{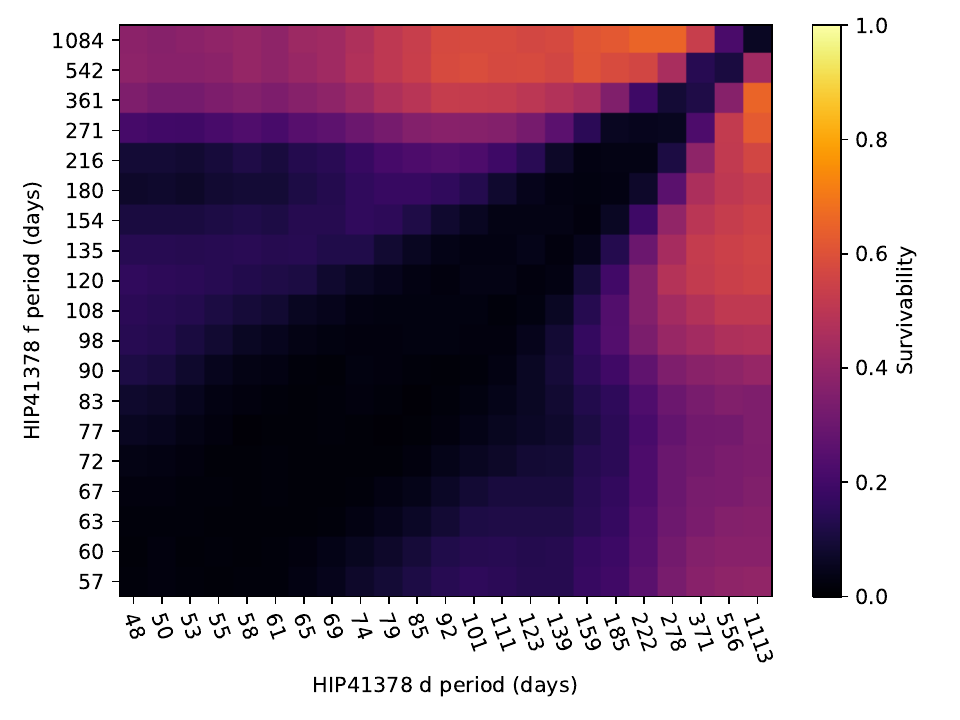}
\caption{\label{fig:period likelihoods}  System stability of the HIP~41378 system for all possible periods of planets~d and~f (see Sec.~\ref{sec:stability}). It is unlikely that both planets have short periods ( $P \lesssim 300$ days), because then their orbits must be highly eccentric and they would interact with planet~c.  Similarly, these two outer planets will be unstable if they both have similar periods. We therefore find  that one planet having a long period and the other having a shorter period is the most likely scenario.}
\end{center}
\end{figure}

For both HIP~41378f and HIP~41378d, we find that $g$, $e$, and $e \sin \omega$ have fairly well-determined values. In contrast, the parameter $\omega$ and combination $e \cos \omega$ are only poorly constrained and so are not listed. Notable is that most possible periods are consistent with non-zero eccentricity at the $> 2 \sigma$, and even the lowest possible eccentricity is $1 \sigma > $ than e = 0, indicating that both planets are most likely on eccentric orbits (see Figure \ref{fig:eccentricites}) .


\begin{deluxetable}{l l l l l}[bt]
\tabletypesize{\scriptsize}
\tablecaption{Photoeccentric Analysis for HIP41378 d\label{tab:photoeccentricd}}
\tablehead{
\colhead{Period } & \colhead{ $g(e,\omega)$ } & \colhead{ $e$ }  & \colhead{ $e \sin \omega$ }
}
\startdata
1114  & $1.47_{-0.138}^{+0.229}$ & $0.517_{-0.16}^{+0.221}$ & $0.262_{-0.242}^{+0.136}$ \\
557  & $1.154_{-0.113}^{+0.171}$ & $0.304_{-0.191}^{+0.296}$ & $0.07_{-0.139}^{+0.135}$ \\
371  & $1.022_{-0.072}^{+0.135}$ & $0.189_{-0.139}^{+0.315}$ & $-0.008_{-0.128}^{+0.102}$ \\
278  & $0.974_{-0.096}^{+0.095}$ & $0.192_{-0.133}^{+0.312}$ & $-0.066_{-0.149}^{+0.099}$ \\
222  & $0.936_{-0.111}^{+0.106}$ & $0.229_{-0.159}^{+0.316}$ & $-0.115_{-0.171}^{+0.128}$ 
\enddata
\end{deluxetable}

\begin{deluxetable}{l l l l l}[bt]
\tabletypesize{\scriptsize}
\tablecaption{Photoeccentric Analysis for HIP41378 f\label{tab:photoeccentricf}}
\tablehead{
\colhead{Period } & \colhead{ $g(e,\omega)$ } & \colhead{ $e$ }  & \colhead{ $e \sin \omega$ }
}
\startdata

1084  & $1.337_{-0.03}^{+0.031}$ & $0.406_{-0.108}^{+0.273}$ & $0.217_{-0.236}^{+0.061}$ \\
542  & $1.059_{-0.024}^{+0.026}$ & $0.153_{-0.089}^{+0.326}$ & $0.035_{-0.104}^{+0.031}$ \\
361  & $0.931_{-0.022}^{+0.023}$ & $0.177_{-0.096}^{+0.343}$ & $-0.095_{-0.11}^{+0.036}$ \\
271  & $0.844_{-0.019}^{+0.02}$ & $0.288_{-0.108}^{+0.316}$ & $-0.2_{-0.126}^{+0.039}$ \\
216  & $0.784_{-0.017}^{+0.019}$ & $0.362_{-0.109}^{+0.29}$ & $-0.274_{-0.13}^{+0.04}$  
\enddata
\end{deluxetable}

\subsection{\textbf{Orbital Overlap}}
\label{sec:stability}
\textbf{By using the results of the photoeccentric analysis, we perform a first-order stability analysis by calculating the possible orbits of planets~d and~f and excluding combinations of parameters where the planets' come within 3.5 mutual Hill radii of one another \citep{kane:2016}, with the mutual Hill radius of two objects given by}

\begin{equation}
r_H = 0.5(a_1 + a_2)[(m_1 + m_2) / M]^{1/3}
\end{equation}

\textbf{where we take $M$ to be the mass of the host star (Table~\ref{tab:stellar}) and $m_1\ /\ a_1$ and $m_2\ /\ a_2$ are the masses / semi-major axes of planets d and f. Since the masses of planets d and f are unknown, we conservatively choose the smallest reasonable masses. We use the publicly available \texttt{Forecaster} code \citep{chen:2017} to estimate the probabilistic masses of the two planets given their radii ($r= 0.33R_J, 0.88 R_J$ for planets d and f respectively). We then take the values one sigma below the median masses as our conservative mass estimate for the planets. We find masses of 0.02 $M_J$ \& 0.2 $M_J$ for planets d and f, which results in a value of $r_H / (a_1 + a_2)$ of $\sim 3$\%.} 

Since the eccentricity and semi-major axis span a wide range of values, we draw samples from the posterior distributions obtained from the MCMC fits discussed in Sec~\ref{sec:k2obs} and in our photoeccentric analysis. Since there are 20 possible periods for planet f and 23 for d, we run an MCMC analysis for each possible period, and perform a stability check for each pair of $20*23$ periods. In this way, we calculate the likelihood for the two planets to have  orbits with overlapping Hill spheres.  In addition to checking for Hill sphere crossings, we also demand that any given orbit of HIP41378~f and d does not overlap with the orbit of HIP 41378~c, which has a well-defined period and semi-major axis. 

An important point is that in the analysis above we do not consider the effects of the fifth planet HIP41378~e. Due to only observing a single transit, we are not able to constrain its period or semi-major axis and so elect to disregard any effects it may have on the system.


\begin{figure}[bt!]
\begin{center}
\includegraphics[width=3.5in,angle=0]{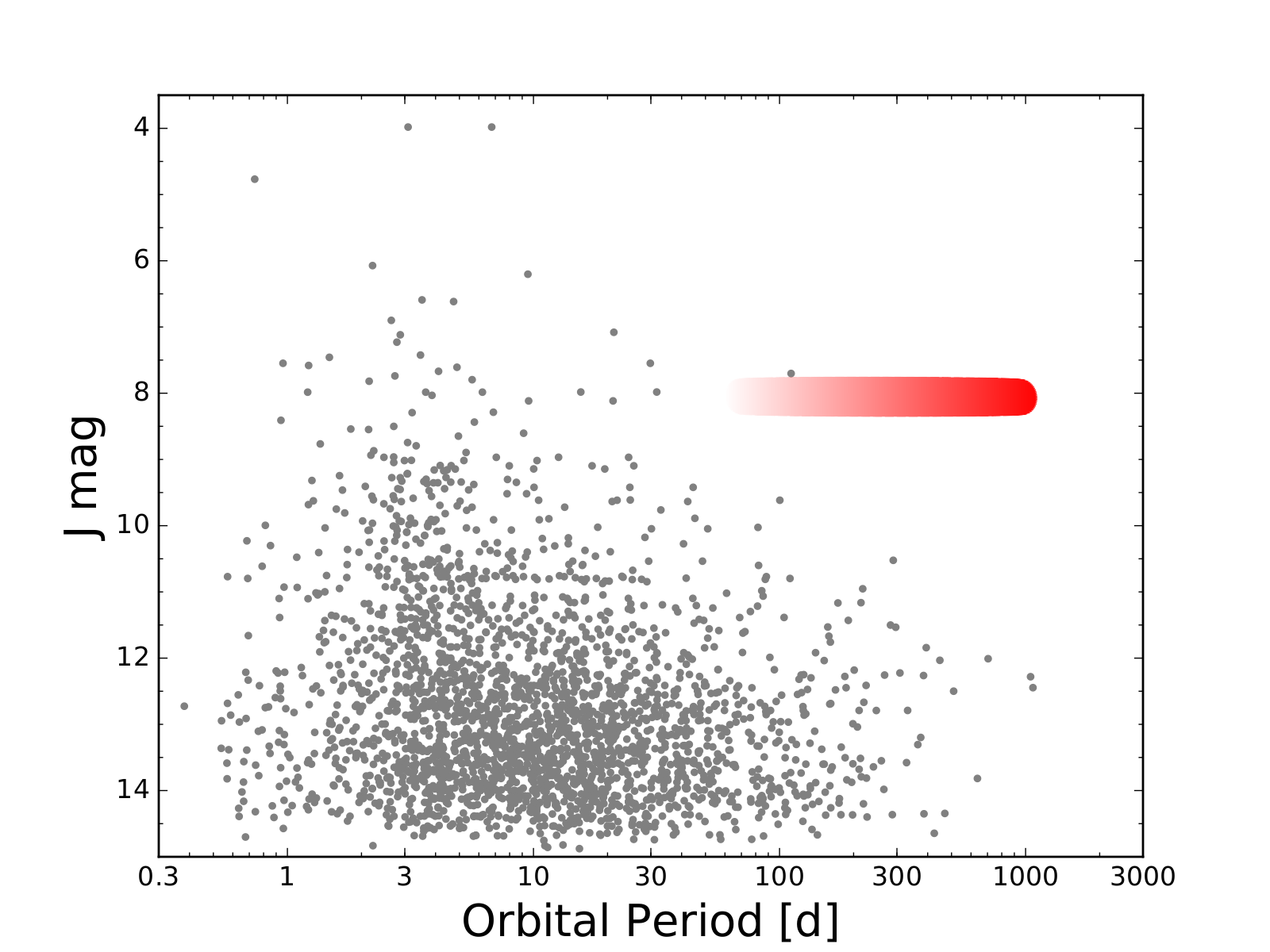}
\caption{\label{fig:per_jmag} HIP~41378d and~f in context: orbital period vs.\ $J$~mag for all known transiting planets. The shaded red lozenge approximately indicates HIP~41378d and~f ---  whatever the orbital periods of these planets, this system is several magnitudes brighter than any comparable systems. \vspace{0.15in} }
\end{center}
\end{figure}

We show the results of this analysis in Figure~\ref{fig:period likelihoods}, with darker colors indicating higher probability of overlap. At low periods (p $\lesssim 300 $ days) there is a higher chance of overlap than long periods.  This is likely due to the fact that at low periods, the photoeccentric analysis predicts increasing eccentricities, shown in Figure~\ref{fig:eccentricites}, making it much more likely that the orbits will overlap with either each other or with planet c. Additionally, the dark diagonal indicates that similar periods for f and d are highly disfavored.

In the most favored scenario (i.e. the one with the highest relative survivability), HIP41378~d has $p=1114$~days and HIP41378~f has $p=361$~or 542~days; Fig.~\ref{fig:eccentricites} shows that this scenario also corresponds to the lowest eccentricities for these two planets.


\section{Discussion}
\label{sec:conclusions}
We analyze new transits of four out of the five planets in the HIP41378 system using \textit{K2} data, two of which previously only had a single observed transit. We study the possible periods of the two planets, and also employ a photoeccentric analysis to study their eccentricity distributions. We find that the eccentricity of the planets increases with decreasing period, however this implies that their orbits will overlap and so disfavors short periods. 

We also observe one additional transit each of planets b and c using \textit{Spitzer} IRAC, providing a sufficient baseline to check for TTVs. For HIP41378 b we find all observations consistent with a linear ephemeris ($t_0 = 2457152.281 \pm{0.015}\ \mathrm{BJD}\  \&\  \mathrm{p} = 15.572119 \pm{0.000022}$ days). For HIP41378 c the \textit{Spitzer} photometry, which occurs roughly at the midpoint between the \textit{K2} campaigns, implies a transit deviation on the order of 50 - 100 minutes. Our \textit{Spitzer} analysis is consistent with two independent, external analyses performed on the same data set. 

\subsection{Follow-up Opportunities}
\label{sec: future obs}
For HIP41378~f, it might seem that such a long-period planet around such a bright star would be an attractive target for high-contrast characterization.  Unfortunately, the  system lies 106~pc away and so even a 1084~d (2.2~AU) orbit places HIP~41378f just 20~mas from its host star.  Assuming a Jupiter-like geometric albedo of 0.35 and a Lambertian phase function, the most favorable planet/star contrasts define a locus from $6\times 10^{-8}$ at 6~mas (for a 217~d period), to $7\times10^{-9}$ at 18~mas (for $P=1084$~d). These values appear to lie just beyond the regime accessible to proposed high-contrast instruments on the next generation of ground-based telescopes \citep{macintosh:2006pfi,beuzit:2006,crossfield:2016b}. Nonetheless, that the planets could come so close to detection bodes well for high-contrast characterization of long-period planets around nearby stars discovered via single transits in \textit{TESS} photometry \citep{villanueva:2018}. Additionally, we find that JWST transmission spectroscopy of the planets is possible at a S/N of $8-10$  for a cloud-free H2-dominated atmospheres if the systematic noise can be kept extremely low ($\sim$ 5 ppm). While this seems like a strict requirement, it is nonetheless interesting to note that such measurements may be feasible for any or all of the larger planets in the system, if their periods can be properly constrained. 

The two outer planets d and f also fall into a less-widely appreciated class of planets, transiting giants on ultra-long periods (T-GULPs).  T-GULPs are those planets with the longest orbital periods, orbiting at the widest separations, and consequently having the lowest known equilibrium temperatures of any known transiting planet. Figure~\ref{fig:per_jmag} lists the known T-GULPs (see also table 7 of \cite{beichman:2018} for a list of the properties of their properties).  Interestingly, few other T-GULPs are known to be in multi-planet systems, and no others orbit stars as bright as HIP~41378 ($V=9$~mag). Whatever their true periods, HIP~41378d and~f (together with their sibling planets) form an exceptional system that will be studied for many years to come.  

The sample of T-GULPs  will likely grow only slowly in the years to come, since \textit{TESS} and other future transit missions will not observe any single field of view nearly as long as \textit{Kepler}.  Only through the extraordinary endurance of \textit{K2} was this observatory able to  redetect the transits of HIP~41378d and~f.  \textit{TESS} will find a few  longer-period planets in its continuous viewing zones \citep[e.g.,][]{sullivan:2015}, but only through an extended mission can the population of truly long-period T-GULPS be substantially expanded.

Because the typical T-GULP has only a few transits observed, the effects of additional perturbing bodies or simple ephemeris drift could eventually result in many of these rare specimens being lost. The situation is even more complicated for HIP~41378d and~f, since only a finite range of possible periods are known. Such long-duration transits (13 hours for planet d and 19 hours for planet f) are challenging to observe from the ground (though it can be done e.g., \cite{shporer:2010}). In contrast, space-based transit photometry is a proven technique for producing high-quality system parameters. We have shown here that \textit{Spitzer} is capable of retrieving transits of the two smaller planets in the system, measuring their transit times to within a few minutes. This implies that it will be easy to observe planets d and f, larger planets with longer transit durations. 

By employing a strategic observing strategy (i.e. observing during the fourth longest period to simultaneously check for the eighth and sixteenth longest periods), and using the mutual likelihood plot of the planet periods (figure \ref{fig:period likelihoods}, it may be possible to pin down the periods of both HIP41378-d and f with only a few additional measurements. We list the future expected transits for the longest periods of each planet in tables \ref{tab:future f} and \ref{tab:future d}.

HIP41378 will also be in the field of view of \textit{TESS} camera 1 in sector 7 (calculated using the Web Tess Viewing tool\footnote{https://heasarc.gsfc.nasa.gov/cgi-bin/tess/webtess/wtv.py}), from 01-07-2019 to 02-04-2019. This time frame lines up with transits of 6 of the possible periods of planet d (53, 55, 58, 74, 111, and 222 days) and 4 of the possible periods of planet f (57, 60, 77, 120 days). This viewing window also coincides with an expected transit of planet c, allowing us to add an additional point to the TTV analysis separated by $\sim$ 200 days from the previous measurement.

\begin{deluxetable*}{l  c c c c c c c c c }[bt]
\tabletypesize{\scriptsize}
\tablecaption{K2 Fit Parameters\label{tab:k2 fit params}}
\tablehead{
\colhead{Planet name} & \colhead{$T_0$ } & \colhead{Period} & \colhead{$a / r_s$} & \colhead{$i$} & \colhead{$r_p / r_s$} & \colhead{q1} & \colhead{q2} \\
\colhead{} & \colhead{ [BJD$_\textrm{TDB} - 2454833$] } & \colhead{ (days)} & \colhead{} & \colhead{degrees} & \colhead{} & \colhead{} & \colhead{} 
}
\startdata
HIP41378 b & 2319.2818$^{+0.0012}_{-0.0012}$ & 15.572098$^{+0.000018}_{-0.000019}$ & 21$^{+2}_{-5}$ & 88.8$^{+0.8}_{-1.4}$ & 0.01843$^{+0.0011}_{-0.00037}$ & 0.463$^{+0.015}_{-0.016}$ & 0.064$^{+0.028}_{-0.028}$\\
HIP41378 c & 2330.1609$^{+0.0023}_{-0.0027}$ & 31.70648$^{+0.00024}_{-0.00019}$ & 22$^{+57}_{-7}$ & 87.5$^{+2.2}_{-1.4}$ & 0.0200$^{+0.018}_{-0.0037}$ & 0.456$^{+0.017}_{-0.016}$ & 0.050$^{+0.030}_{-0.027}$\\
HIP41378 d & 2333.2604$^{+0.0017}_{-0.0017}$ & 1113.4491$^{+0.0018}_{-0.0018}$ & 533$^{+81}_{-56}$ & 89.930$^{+0.025}_{-0.018}$ & 0.02560$^{+0.0005}_{-0.0007}$ & 0.444$^{+0.015}_{-0.014}$ & 0.028$^{+0.022}_{-0.018}$\\
HIP41378 e & 2309.0194$^{+0.001}_{-0.001}$ & - & 283$^{+172}_{-177}$ & 89.910$^{+0.22}_{-0.045}$ & 0.03686$^{+0.0011}_{-0.0008}$ & 0.451$^{+0.015}_{-0.015}$ & 0.041$^{+0.026}_{-0.024}$\\
HIP41378 f & 2353.91423$^{+0.00039}_{-0.00038}$ & 1084.16156$^{+0.00040}_{-0.00042}$ & 460$^{+5}_{-4}$ & 89.98$^{+0.009}_{-0.006}$ & 0.06602$^{+0.00017}_{-0.00016}$ & 0.455$^{+0.015}_{-0.014}$ & 0.044$^{+0.029}_{-0.026}$
\enddata
\tablecomments{\textbf{The limb darkening parameters for planets b-e use the posterior values from the fit for planet f as gaussian priors.}}
\end{deluxetable*}

\begin{deluxetable*}{l  c c c c c c }[bt]
\tabletypesize{\scriptsize}
\tablecaption{Spitzer Fit Parameters\label{tab:spitzerfitparams}}
\tablehead{
\colhead{Planet name} & \colhead{$T_0$ } & \colhead{Transit Depth } & \colhead{$r_p / r_s$} & \colhead{$a / r_s$} & \colhead{$i$}\\
\colhead{} & \colhead{ [BJD$_\textrm{TDB}$] } & \colhead{ (ppm)} & \colhead{} & \colhead{} & \colhead{degrees}
}
\startdata
HIP41378 b & $2457790.734^{+0.016}_{-0.0035}$ & $374^{+60}_{-65}$ & $0.0194^{+0.0015}_{-0.0016}$ & $22^{+3}_{-7}$ & $89.05^{+0.6}_{-1.3}$\\
HIP41378 c & $2457606.985^{+0.0036}_{-0.0036}$ & $444^{+92}_{-95}$ & $0.0211^{+0.0022}_{-0.0022}$ & $85^{+14}_{-31}$ & $89.6^{+0.2}_{-0.6}$
\enddata
\end{deluxetable*}

\begin{deluxetable*}{c l c c c}[bt]
\tabletypesize{\scriptsize}
\tablecaption{Individual Transit Centers for planet b\label{tab:ttv times b}}
\tablehead{
\colhead{Epoch} & \colhead{Observed} & \colhead{Calculated} & O - C  & \colhead{Data Set} \\
\colhead{} & \colhead{[BJD$_\textrm{TDB} - 2454833$]} & \colhead{ [BJD$_\textrm{TDB} - 2454833$] } & \colhead{ (minutes)} & \colhead{}
}
\startdata
0 & $2319.2797^{+0.0008}_{-0.0006}$ & 2319.2798 & $-0.1^{+1.2}_{0.9}$ & \textit{K2} C5\\
1 & $2334.8499^{+0.0061}_{-0.001}$ & 2334.8519 & $-2.9^{+8.7}_{1.5}$ & \textit{K2} C5\\
2 & $2350.4187^{+0.0055}_{-0.0011}$ & 2350.424 & $-7.7^{+8.0}_{1.6}$ & \textit{K2} C5\\
3 & $2366.0033^{+0.001}_{-0.0039}$ & 2365.9962 & $10.3^{+1.5}_{5.6}$ & \textit{K2} C5\\
41 & $2957.734^{+0.016}_{-0.003}$ & 2957.737 & $-4.0^{+23.2}_{4.3}$ & \textit{Spitzer} \\
71 & $3424.90086^{+0.00021}_{-0.00021}$ & 3424.9007 & $0.2^{+0.3}_{0.3}$ & \textit{K2} C18\\
72 & $3440.47139^{+0.00021}_{-0.00021}$ & 3440.47282 & $-2.1^{+0.3}_{0.3}$ & \textit{K2} C18
\enddata
\tablecomments{The calculated ephemeris is given by t = 2319.27979 + (15.57213)$\times$E, where E is the epoch of the transit. Errors on the calculated ephemeris are included in the errors of O-C listed above.}
\end{deluxetable*}

\begin{deluxetable*}{c l c c c}[bt]
\tabletypesize{\scriptsize}
\tablecaption{Individual Transit Centers for planet c\label{tab:ttv times c}}
\tablehead{
\colhead{Epoch} & \colhead{Observed} & \colhead{Calculated} & O - C  & \colhead{Data Set} \\
\colhead{} & \colhead{[BJD$_\textrm{TDB} - 2454833$]} & \colhead{ [BJD$_\textrm{TDB} - 2454833$] } & \colhead{ (minutes)} & \colhead{}
}
\startdata
0 & $2330.16576^{+0.00245}_{-0.00272}$ & 2330.15167 & $20.3^{+3.5}_{3.9}$ & \textit{K2} C5\\
1 & $2361.86375^{+0.00297}_{-0.00284}$ & 2361.85836 & $7.8^{+4.3}_{4.1}$ & \textit{K2} C5\\
14 & $2773.986^{+0.0036}_{-0.0036}$ & 2774.04529 & $-85.4^{+5.2}_{5.2}$ & \textit{Spitzer}\\
35 & $3439.88676^{+0.0008}_{-0.00074}$ & 3439.8857 & $1.5^{+1.2}_{1.1}$ & \textit{K2} C18 
\enddata
\tablecomments{The calculated ephemeris is given by t = 2330.15160 + (31.70669)$\times$E, where E is the epoch of the transit. Errors on the calculated ephemeris are included in the errors of O-C listed above.}
\end{deluxetable*}

\begin{deluxetable*}{l l l l l}[bt]
\tabletypesize{\scriptsize}
\tablecaption{Future Transit Windows for HIP 41378 d\label{tab:future d}}
\tablehead{
\colhead{Period } & \colhead{$T_0$ } & \colhead{ Start} & \colhead{Midpoint} & \colhead{End} \\
\colhead{[d] } & \colhead{ [BJD$_\textrm{TDB} - 2454833$] } & \colhead{ [UT]} & \colhead{[UT]} & \colhead{[UT]} 
}
1113.45 & 4560.1586 $^{+0.0022}_{-0.0022}$ & 2021-06-27 09:28:32 & 2021-06-27 15:48:21 & 2021-06-27 22:07:47 \\
\hline
556.72 & 4003.434 $^{+0.0014}_{-0.0014}$ & 2019-12-18 16:05:13 & 2019-12-18 22:24:53 & 2019-12-19 04:44:23 \\
\hline
371.15 & 3817.8591 $^{+0.0011}_{-0.0011}$ & 2019-06-16 02:17:27 & 2019-06-16 08:37:09 & 2019-06-16 14:56:33 \\
\hline
278.36 & 3725.0717 $^{+0.001}_{-0.001}$ & 2019-03-15 07:23:12 & 2019-03-15 13:43:11 & 2019-03-15 20:02:55 \\
\hline
222.69 & 3669.3992 $^{+0.0009}_{-0.0009}$ & 2019-01-18 15:15:19 & 2019-01-18 21:34:53 & 2019-01-19 03:54:17 \\
\hline
185.57 & 3632.2843 $^{+0.0009}_{-0.0009}$ & 2018-12-12 12:29:47 & 2018-12-12 18:49:21 & 2018-12-13 01:08:38 \\
\hline
159.06 & 3605.7736 $^{+0.0008}_{-0.0009}$ & 2018-11-16 00:14:43 & 2018-11-16 06:34:01 & 2018-11-16 12:53:01 \\
159.06 & 3764.8378 $^{+0.001}_{-0.001}$ & 2019-04-24 01:47:08 & 2019-04-24 08:06:26 & 2019-04-24 14:25:25 \\
\hline
139.18 & 3585.8906 $^{+0.0009}_{-0.0009}$ & 2018-10-27 03:02:44 & 2018-10-27 09:22:24 & 2018-10-27 15:41:57 \\
139.18 & 3725.0717 $^{+0.001}_{-0.001}$ & 2019-03-15 07:23:35 & 2019-03-15 13:43:14 & 2019-03-15 20:02:47 \\
\hline
123.72 & 3570.426 $^{+0.0009}_{-0.0008}$ & 2018-10-11 15:53:31 & 2018-10-11 22:13:24 & 2018-10-12 04:33:04 \\
123.72 & 3694.1425 $^{+0.001}_{-0.001}$ & 2019-02-12 09:05:23 & 2019-02-12 15:25:15 & 2019-02-12 21:44:56 \\
\hline
111.34 & 3558.0544 $^{+0.0009}_{-0.0008}$ & 2018-09-29 06:58:25 & 2018-09-29 13:18:16 & 2018-09-29 19:37:49 \\
111.34 & 3669.3993 $^{+0.001}_{-0.0009}$ & 2019-01-18 15:15:06 & 2019-01-18 21:34:56 & 2019-01-19 03:54:29 \\
111.34 & 3780.7442 $^{+0.0011}_{-0.0011}$ & 2019-05-09 23:31:46 & 2019-05-10 05:51:37 & 2019-05-10 12:11:09 \\
\hline
101.22 & 3649.1547 $^{+0.0009}_{-0.0009}$ & 2018-12-29 09:23:17 & 2018-12-29 15:42:46 & 2018-12-29 22:02:04 \\
101.22 & 3750.3774 $^{+0.001}_{-0.001}$ & 2019-04-09 14:43:54 & 2019-04-09 21:03:23 & 2019-04-10 03:22:40 \\
\hline
92.79 & 3632.2843 $^{+0.0009}_{-0.0009}$ & 2018-12-12 12:30:11 & 2018-12-12 18:49:22 & 2018-12-13 01:08:15 \\
92.79 & 3725.0717 $^{+0.001}_{-0.001}$ & 2019-03-15 07:24:05 & 2019-03-15 13:43:15 & 2019-03-15 20:02:07 \\
\hline
85.65 & 3618.0093 $^{+0.0009}_{-0.0009}$ & 2018-11-28 05:53:57 & 2018-11-28 12:13:25 & 2018-11-28 18:32:42 \\
85.65 & 3703.6593 $^{+0.001}_{-0.001}$ & 2019-02-21 21:29:52 & 2019-02-22 03:49:19 & 2019-02-22 10:08:36 \\
85.65 & 3789.3092 $^{+0.0011}_{-0.0011}$ & 2019-05-18 13:05:47 & 2019-05-18 19:25:13 & 2019-05-19 01:44:29 \\
\hline
79.53 & 3605.7736 $^{+0.0009}_{-0.0009}$ & 2018-11-16 00:14:24 & 2018-11-16 06:33:59 & 2018-11-16 12:53:22 \\
79.53 & 3685.3057 $^{+0.0009}_{-0.001}$ & 2019-02-03 13:00:36 & 2019-02-03 19:20:11 & 2019-02-04 01:39:34 \\
79.53 & 3764.8378 $^{+0.001}_{-0.0011}$ & 2019-04-24 01:46:49 & 2019-04-24 08:06:23 & 2019-04-24 14:25:45 \\
\hline
74.23 & 3595.1693 $^{+0.0008}_{-0.0008}$ & 2018-11-05 09:44:06 & 2018-11-05 16:03:47 & 2018-11-05 22:23:16 \\
74.23 & 3669.3992 $^{+0.0009}_{-0.0009}$ & 2019-01-18 15:15:13 & 2019-01-18 21:34:53 & 2019-01-19 03:54:23 \\
74.23 & 3743.6292 $^{+0.001}_{-0.001}$ & 2019-04-02 20:46:20 & 2019-04-03 03:06:00 & 2019-04-03 09:25:30 \\
\hline
69.59 & 3585.8906 $^{+0.0009}_{-0.0009}$ & 2018-10-27 03:02:47 & 2018-10-27 09:22:25 & 2018-10-27 15:41:48 \\
69.59 & 3655.4811 $^{+0.0009}_{-0.0009}$ & 2019-01-04 17:13:13 & 2019-01-04 23:32:50 & 2019-01-05 05:52:14 \\
69.59 & 3725.0717 $^{+0.001}_{-0.001}$ & 2019-03-15 07:23:39 & 2019-03-15 13:43:15 & 2019-03-15 20:02:39 \\
69.59 & 3794.6623 $^{+0.0011}_{-0.0011}$ & 2019-05-23 21:34:04 & 2019-05-24 03:53:40 & 2019-05-24 10:13:04 \\
\hline
65.5 & 3577.7034 $^{+0.0008}_{-0.0008}$ & 2018-10-18 22:33:18 & 2018-10-19 04:52:57 & 2018-10-19 11:12:24 \\
65.5 & 3643.2005 $^{+0.0009}_{-0.0009}$ & 2018-12-23 10:29:00 & 2018-12-23 16:48:39 & 2018-12-23 23:08:05 \\
65.5 & 3708.6975 $^{+0.001}_{-0.001}$ & 2019-02-26 22:24:42 & 2019-02-27 04:44:20 & 2019-02-27 11:03:46 \\
65.5 & 3774.1945 $^{+0.001}_{-0.0011}$ & 2019-05-03 10:20:23 & 2019-05-03 16:40:01 & 2019-05-03 22:59:27 \\
\hline
61.86 & 3570.426 $^{+0.0008}_{-0.0008}$ & 2018-10-11 15:53:42 & 2018-10-11 22:13:23 & 2018-10-12 04:32:53 \\
61.86 & 3632.2843 $^{+0.0009}_{-0.0009}$ & 2018-12-12 12:29:38 & 2018-12-12 18:49:19 & 2018-12-13 01:08:49 \\
61.86 & 3694.1425 $^{+0.001}_{-0.001}$ & 2019-02-12 09:05:34 & 2019-02-12 15:25:14 & 2019-02-12 21:44:45 \\
61.86 & 3756.0008 $^{+0.001}_{-0.001}$ & 2019-04-15 05:41:31 & 2019-04-15 12:01:10 & 2019-04-15 18:20:40 \\
\hline
58.6 & 3563.9146 $^{+0.001}_{-0.0009}$ & 2018-10-05 03:37:27 & 2018-10-05 09:56:58 & 2018-10-05 16:16:26 \\
58.6 & 3622.5171 $^{+0.001}_{-0.001}$ & 2018-12-02 18:05:11 & 2018-12-03 00:24:41 & 2018-12-03 06:44:10 \\
58.6 & 3681.1197 $^{+0.0011}_{-0.001}$ & 2019-01-30 08:32:54 & 2019-01-30 14:52:24 & 2019-01-30 21:11:53 \\
58.6 & 3739.7223 $^{+0.0012}_{-0.0011}$ & 2019-03-29 23:00:38 & 2019-03-30 05:20:07 & 2019-03-30 11:39:37 \\
\hline
55.67 & 3558.0544 $^{+0.0008}_{-0.0008}$ & 2018-09-29 06:58:51 & 2018-09-29 13:18:16 & 2018-09-29 19:37:25 \\
55.67 & 3613.7268 $^{+0.0009}_{-0.0009}$ & 2018-11-23 23:07:11 & 2018-11-24 05:26:37 & 2018-11-24 11:45:45 \\
55.67 & 3669.3993 $^{+0.0009}_{-0.0009}$ & 2019-01-18 15:15:32 & 2019-01-18 21:34:57 & 2019-01-19 03:54:05 \\
55.67 & 3725.0717 $^{+0.001}_{-0.001}$ & 2019-03-15 07:23:52 & 2019-03-15 13:43:17 & 2019-03-15 20:02:25 \\
\hline
53.02 & 3605.7736 $^{+0.0009}_{-0.0009}$ & 2018-11-16 00:14:13 & 2018-11-16 06:33:56 & 2018-11-16 12:53:25 \\
53.02 & 3658.795 $^{+0.0009}_{-0.0009}$ & 2019-01-08 00:45:00 & 2019-01-08 07:04:43 & 2019-01-08 13:24:12 \\
53.02 & 3711.8163 $^{+0.001}_{-0.001}$ & 2019-03-02 01:15:48 & 2019-03-02 07:35:31 & 2019-03-02 13:55:00 \\
53.02 & 3764.8377 $^{+0.001}_{-0.0011}$ & 2019-04-24 01:46:36 & 2019-04-24 08:06:19 & 2019-04-24 14:25:48 \\
\hline
50.61 & 3598.5433 $^{+0.0009}_{-0.0011}$ & 2018-11-08 18:42:30 & 2018-11-09 01:02:20 & 2018-11-09 07:21:29 \\
50.61 & 3649.1546 $^{+0.001}_{-0.0011}$ & 2018-12-29 09:22:48 & 2018-12-29 15:42:38 & 2018-12-29 22:01:47 \\
50.61 & 3699.7659 $^{+0.001}_{-0.0012}$ & 2019-02-18 00:03:07 & 2019-02-18 06:22:56 & 2019-02-18 12:42:05 \\
50.61 & 3750.3773 $^{+0.0011}_{-0.0013}$ & 2019-04-09 14:43:25 & 2019-04-09 21:03:14 & 2019-04-10 03:22:23 \\
\hline
48.41 & 3591.9419 $^{+0.0009}_{-0.0009}$ & 2018-11-02 04:16:48 & 2018-11-02 10:36:22 & 2018-11-02 16:55:41 \\
48.41 & 3640.3528 $^{+0.0009}_{-0.0009}$ & 2018-12-20 14:08:25 & 2018-12-20 20:27:58 & 2018-12-21 02:47:17 \\
48.41 & 3688.7636 $^{+0.001}_{-0.001}$ & 2019-02-07 00:00:01 & 2019-02-07 06:19:34 & 2019-02-07 12:38:53 \\
48.41 & 3737.1744 $^{+0.001}_{-0.001}$ & 2019-03-27 09:51:37 & 2019-03-27 16:11:10 & 2019-03-27 22:30:29 
\enddata
\end{deluxetable*}

\begin{deluxetable*}{l l l l l}[bt]
\tabletypesize{\scriptsize}
\tablecaption{Future Transit Windows of HIP41378 f\label{tab:future f}}
\tablehead{
\colhead{Period } & \colhead{$T_0$ } & \colhead{ Start} & \colhead{Midpoint} & \colhead{End} \\
\colhead{[d] } & \colhead{ [BJD$_\textrm{TDB} - 2454833$] } & \colhead{ [UT]} & \colhead{[UT]} & \colhead{[UT]} 
}
1084.16 & 4522.23776 $^{+0.00048}_{-0.00047}$ & 2021-05-20 08:16:43 & 2021-05-20 17:42:22 & 2021-05-21 03:08:00 \\
\hline
542.08 & 3980.15685 $^{+0.00029}_{-0.0003}$ & 2019-11-25 06:20:14 & 2019-11-25 15:45:52 & 2019-11-26 01:11:29 \\
\hline
361.39 & 3799.46322 $^{+0.00024}_{-0.00024}$ & 2019-05-28 13:41:25 & 2019-05-28 23:07:02 & 2019-05-29 08:32:38 \\
\hline
271.04 & 3709.1164 $^{+0.00022}_{-0.00022}$ & 2019-02-27 05:21:59 & 2019-02-27 14:47:36 & 2019-02-28 00:13:14 \\
\hline
216.83 & 3654.90831 $^{+0.0002}_{-0.0002}$ & 2019-01-04 00:22:21 & 2019-01-04 09:47:58 & 2019-01-04 19:13:34 \\
\hline
180.69 & 3618.76958 $^{+0.00019}_{-0.00019}$ & 2018-11-28 21:02:35 & 2018-11-29 06:28:11 & 2018-11-29 15:53:48 \\
180.69 & 3799.46321 $^{+0.00024}_{-0.00024}$ & 2019-05-28 13:41:24 & 2019-05-28 23:07:01 & 2019-05-29 08:32:38 \\
\hline
154.88 & 3592.9562 $^{+0.00019}_{-0.00018}$ & 2018-11-03 01:31:21 & 2018-11-03 10:56:55 & 2018-11-03 20:22:30 \\
154.88 & 3747.83646 $^{+0.00023}_{-0.00022}$ & 2019-04-06 22:38:54 & 2019-04-07 08:04:29 & 2019-04-07 17:30:04 \\
\hline
135.52 & 3573.59617 $^{+0.00019}_{-0.00018}$ & 2018-10-14 16:52:51 & 2018-10-15 02:18:29 & 2018-10-15 11:44:07 \\
135.52 & 3709.1164 $^{+0.00022}_{-0.00021}$ & 2019-02-27 05:21:59 & 2019-02-27 14:47:36 & 2019-02-28 00:13:14 \\
\hline
120.46 & 3558.53838 $^{+0.00018}_{-0.00017}$ & 2018-09-29 15:29:40 & 2018-09-30 00:55:15 & 2018-09-30 10:20:52 \\
120.46 & 3679.0008 $^{+0.00021}_{-0.0002}$ & 2019-01-28 02:35:33 & 2019-01-28 12:01:09 & 2019-01-28 21:26:46 \\
120.46 & 3799.46323 $^{+0.00024}_{-0.00023}$ & 2019-05-28 13:41:27 & 2019-05-28 23:07:02 & 2019-05-29 08:32:39 \\
\hline
108.42 & 3654.90831 $^{+0.0002}_{-0.0002}$ & 2019-01-04 00:22:20 & 2019-01-04 09:47:57 & 2019-01-04 19:13:35 \\
108.42 & 3763.32449 $^{+0.00023}_{-0.00023}$ & 2019-04-22 10:21:38 & 2019-04-22 19:47:15 & 2019-04-23 05:12:52 \\
\hline
98.56 & 3635.19628 $^{+0.0002}_{-0.00019}$ & 2018-12-15 07:17:01 & 2018-12-15 16:42:38 & 2018-12-16 02:08:15 \\
98.56 & 3733.75644 $^{+0.00022}_{-0.00022}$ & 2019-03-23 20:43:39 & 2019-03-24 06:09:16 & 2019-03-24 15:34:53 \\
\hline
90.35 & 3618.76958 $^{+0.00019}_{-0.00019}$ & 2018-11-28 21:02:37 & 2018-11-29 06:28:11 & 2018-11-29 15:53:46 \\
90.35 & 3709.1164 $^{+0.00022}_{-0.00021}$ & 2019-02-27 05:22:02 & 2019-02-27 14:47:37 & 2019-02-28 00:13:11 \\
90.35 & 3799.46322 $^{+0.00024}_{-0.00024}$ & 2019-05-28 13:41:27 & 2019-05-28 23:07:02 & 2019-05-29 08:32:36 \\
\hline
83.4 & 3604.87008 $^{+0.00019}_{-0.00019}$ & 2018-11-14 23:27:17 & 2018-11-15 08:52:54 & 2018-11-15 18:18:31 \\
83.4 & 3688.26714 $^{+0.00021}_{-0.00021}$ & 2019-02-06 08:59:03 & 2019-02-06 18:24:40 & 2019-02-07 03:50:17 \\
83.4 & 3771.6642 $^{+0.00023}_{-0.00023}$ & 2019-04-30 18:30:49 & 2019-05-01 03:56:26 & 2019-05-01 13:22:04 \\
\hline
77.44 & 3592.95633 $^{+95.31082}_{-0.00026}$ & 2018-11-03 01:31:41 & 2018-11-03 10:57:07 & 2018-11-03 20:22:55 \\
77.44 & 3670.39648 $^{+101.26775}_{-0.00029}$ & 2019-01-19 12:05:28 & 2019-01-19 21:30:55 & 2019-01-20 06:56:42 \\
77.44 & 3747.83662 $^{+107.22467}_{-0.00031}$ & 2019-04-06 22:39:16 & 2019-04-07 08:04:43 & 2019-04-07 17:30:30 \\
\hline
72.28 & 3582.63095 $^{+87.76528}_{-0.00023}$ & 2018-10-23 17:43:07 & 2018-10-24 03:08:34 & 2018-10-24 12:34:21 \\
72.28 & 3654.90842 $^{+92.92794}_{-0.00026}$ & 2019-01-04 00:22:39 & 2019-01-04 09:48:07 & 2019-01-04 19:13:54 \\
72.28 & 3727.18588 $^{+98.09059}_{-0.00028}$ & 2019-03-17 07:02:12 & 2019-03-17 16:27:40 & 2019-03-18 01:53:26 \\
72.28 & 3799.46335 $^{+103.25324}_{-0.00031}$ & 2019-05-28 13:41:44 & 2019-05-28 23:07:13 & 2019-05-29 08:32:59 \\
\hline
67.76 & 3573.59617 $^{+0.00019}_{-0.0002}$ & 2018-10-14 16:52:51 & 2018-10-15 02:18:29 & 2018-10-15 11:44:05 \\
67.76 & 3641.35628 $^{+0.0002}_{-0.00021}$ & 2018-12-21 11:07:24 & 2018-12-21 20:33:02 & 2018-12-22 05:58:39 \\
67.76 & 3709.11639 $^{+0.00022}_{-0.00023}$ & 2019-02-27 05:21:58 & 2019-02-27 14:47:36 & 2019-02-28 00:13:12 \\
67.76 & 3776.87651 $^{+0.00024}_{-0.00025}$ & 2019-05-05 23:36:32 & 2019-05-06 09:02:10 & 2019-05-06 18:27:46 \\
\hline
63.77 & 3565.6244 $^{+0.00018}_{-0.00018}$ & 2018-10-06 17:33:30 & 2018-10-07 02:59:08 & 2018-10-07 12:24:45 \\
63.77 & 3629.39862 $^{+0.00019}_{-0.00019}$ & 2018-12-09 12:08:23 & 2018-12-09 21:34:00 & 2018-12-10 06:59:38 \\
63.77 & 3693.17285 $^{+0.00021}_{-0.00021}$ & 2019-02-11 06:43:16 & 2019-02-11 16:08:53 & 2019-02-12 01:34:31 \\
63.77 & 3756.94707 $^{+0.00023}_{-0.00022}$ & 2019-04-16 01:18:09 & 2019-04-16 10:43:46 & 2019-04-16 20:09:24 \\
\hline
60.23 & 3558.53829 $^{+0.00022}_{-63.40115}$ & 2018-09-29 15:29:24 & 2018-09-30 00:55:08 & 2018-09-30 10:20:38 \\
60.23 & 3618.7695 $^{+0.00024}_{-66.5712}$ & 2018-11-28 21:02:20 & 2018-11-29 06:28:04 & 2018-11-29 15:53:34 \\
60.23 & 3679.00071 $^{+0.00025}_{-69.74124}$ & 2019-01-28 02:35:17 & 2019-01-28 12:01:00 & 2019-01-28 21:26:31 \\
60.23 & 3739.23191 $^{+0.00027}_{-72.91127}$ & 2019-03-29 08:08:13 & 2019-03-29 17:33:57 & 2019-03-30 02:59:27 \\
60.23 & 3799.46311 $^{+0.00029}_{-76.08131}$ & 2019-05-28 13:41:09 & 2019-05-28 23:06:53 & 2019-05-29 08:32:23 \\
\hline
57.06 & 3609.25939 $^{+0.00019}_{-0.00019}$ & 2018-11-19 08:47:54 & 2018-11-19 18:13:31 & 2018-11-20 03:39:08 \\
57.06 & 3666.32054 $^{+0.0002}_{-0.0002}$ & 2019-01-15 10:15:57 & 2019-01-15 19:41:34 & 2019-01-16 05:07:11 \\
57.06 & 3723.38169 $^{+0.00022}_{-0.00021}$ & 2019-03-13 11:44:01 & 2019-03-13 21:09:38 & 2019-03-14 06:35:14 \\
57.06 & 3780.44284 $^{+0.00023}_{-0.00023}$ & 2019-05-09 13:12:04 & 2019-05-09 22:37:41 & 2019-05-10 08:03:17 

\enddata
\end{deluxetable*}

\section*{Acknowledgments}
The authors would like to direct the reader to Becker et al.\ (submitted), which also presents an updated analysis of the HIP~41378 system. 

This work is based in part on observations made with the \textit{Spitzer} Space Telescope, which is operated by the Jet Propulsion Laboratory, California Institute of Technology under a contract with NASA. Support for this work was provided by NASA through an award issued by JPL/Caltech.  DB and IJMC acknowledge support from NSF AAG grant 1616648, DB acknowledges support from an NSERC PGS-D scholarship, and IJMC acknowledges support from NASA \textit{K2} GO grants NNH15ZDA001N-15-K2GO4\_2-0018 and NNH16ZDA001N-16-K2GO5\_2-0005.

The authors wish to recognize and acknowledge the very
significant cultural role and reverence that the summit of Maunakea
has always had within the indigenous Hawaiian community.
We are most fortunate to have the opportunity to conduct
observations from this mountain.

{\it Facility:} \facility{Kepler}, \facility{K2}, \facility{Spitzer}, \facility{Keck~I}, \facility{Gaia}

\bibliography{ms}

\end{document}